\documentclass[conference,10pt]{IEEEtran}
\IEEEoverridecommandlockouts
\usepackage{cite}
\usepackage{amsmath,amssymb,amsfonts}
\usepackage{algorithmic}
\usepackage{graphicx}
\usepackage{textcomp}
\usepackage{xcolor}

\usepackage{adjustbox}
\usepackage{booktabs} 
\usepackage{enumitem}
\usepackage{makecell}

\setitemize{leftmargin=*}
\setenumerate{leftmargin=*}

\usepackage[inkscapelatex=false]{svg}
\def\BibTeX{{\rm B\kern-.05em{\sc i\kern-.025em b}\kern-.08em
    T\kern-.1667em\lower.7ex\hbox{E}\kern-.125emX}}

\usepackage{titlesec}
\titlespacing\section{0pt}{0.6\baselineskip}{0.3\baselineskip}
\titlespacing\subsection{0pt}{0.3\baselineskip}{0.15\baselineskip}
\titlespacing\subsubsection{0pt}{0.2\baselineskip}{0.1\baselineskip}

\usepackage{fancyhdr,lipsum}
\fancypagestyle{firstpage}{
  \fancyhf{}
  \fancyhead[C]{To appear at the 2024 IEEE International Conference on Quantum Computing and Engineering (QCE24), September 2024.}
  \fancyfoot[C]{\thepage}
}

\begin{document}

\bstctlcite{IEEEexample:BSTcontrol} 

\title{PO-QA: A Framework for Portfolio Optimization using Quantum Algorithms
\vspace{-5pt}
}

\author{\IEEEauthorblockN{Kamila Zaman\textsuperscript{1,2}, Alberto Marchisio\textsuperscript{1,2}, Muhammad Kashif\textsuperscript{1,2}, and Muhammad Shafique\textsuperscript{1,2}}
\IEEEauthorblockA{\textsuperscript{1}eBrain Lab, Division of Engineering, New York University Abu Dhabi, UAE\\
\textsuperscript{2}Center for Quantum and Topological Systems (CQTS), NYUAD Research Institute, New York University Abu Dhabi, UAE\\
(e-mail: kz2137@nyu.edu, alberto.marchisio@nyu.edu, muhammadkashif@nyu.edu, muhammad.shafique@nyu.edu)
}
\vspace{-20pt}
}


\maketitle
\thispagestyle{firstpage}

\begin{abstract}


Portfolio Optimization (PO) is a financial problem aiming to maximize the net gains while minimizing the risks in a given investment portfolio. The novelty of Quantum algorithms lies in their acclaimed potential and capability to solve complex problems given the underlying Quantum Computing (QC) infrastructure. Utilizing QC's applicable strengths to the finance industry's problems, such as PO, allows us to solve these problems using quantum-based algorithms such as Variational Quantum Eigensolver (VQE) and Quantum Approximate Optimization Algorithm (QAOA). While the Quantum potential for finance is highly impactful, the architecture and composition of the quantum circuits have not yet been properly defined as robust financial frameworks/algorithms as state of the art in present literature for research and design development purposes. In this work, we propose a novel scalable framework, denoted PO-QA, to systematically investigate the variation of quantum parameters (such as rotation blocks, repetitions, and entanglement types) to observe their subtle effect on the overall performance. In our paper, the performance is measured and dictated by convergence to similar ground-state energy values for resultant optimal solutions by each algorithm variation set for QAOA and VQE to the exact eigensolver (classical solution). 
Our results provide effective insights into comprehending PO from the lens of Quantum Machine Learning in terms of convergence to the classical solution, which is used as a benchmark. This study paves the way for identifying efficient configurations of quantum circuits for solving PO and unveiling their inherent inter-relationships. 
\end{abstract}





\section{Introduction and Motivation}

As a fundamental task in the financial field, Portfolio Optimization (PO)~\cite{sen2023portfoliostudy} aims at finding optimal trades, given a budget and a set of assets, for minimizing the risk while maximizing the return. 
According to the Modern Portfolio Theory (MPT)~\cite{portfolioselection_markowitz_1952}, an optimal portfolio profile that balances risks and rewards usually requires high degrees of diversification (i.e., investments in multiple and heterogeneous assets). Solving PO problems with diversified portfolio profiles requires extensive use of quantitative analysis, which dramatically increases the complexity of the system.

The recent advancements and computational capabilities of the current Noise Intermediate-Scale Quantum (NISQ) devices~\cite{Preskill_2018arxiv_QC_NISQ_era, zaman2023survey}, have shown that PO problems can be efficiently solved in the quantum domain using quantum computers~\cite{buonaiuto2023best, abbas2023quantum, s2023potentialquantumtechniquesstock}. 
To implement PO on NISQ devices, we need to process the financial market data using quadratic programming~\cite{Kerenidis_2019AFT_QuantumPortfolio} and solve the equivalent problem of minimizing the ground state energy of the Hamiltonian~\cite{Claudino_2020FrontChem_BenchmarkingVQE}. 
Among different optimization algorithms for PO, frequently used state-of-the-art quantum algorithms are VQE~\cite{Peruzzo_2014_VQE} and QAOA~\cite{Farhi_2014_QAOA}. However, designing efficient architectures for the VQE and QAOA quantum circuits remains an open research problem.

\subsection{Motivational Case Study}

\begin{figure}[ht]
\includegraphics[width=\linewidth]{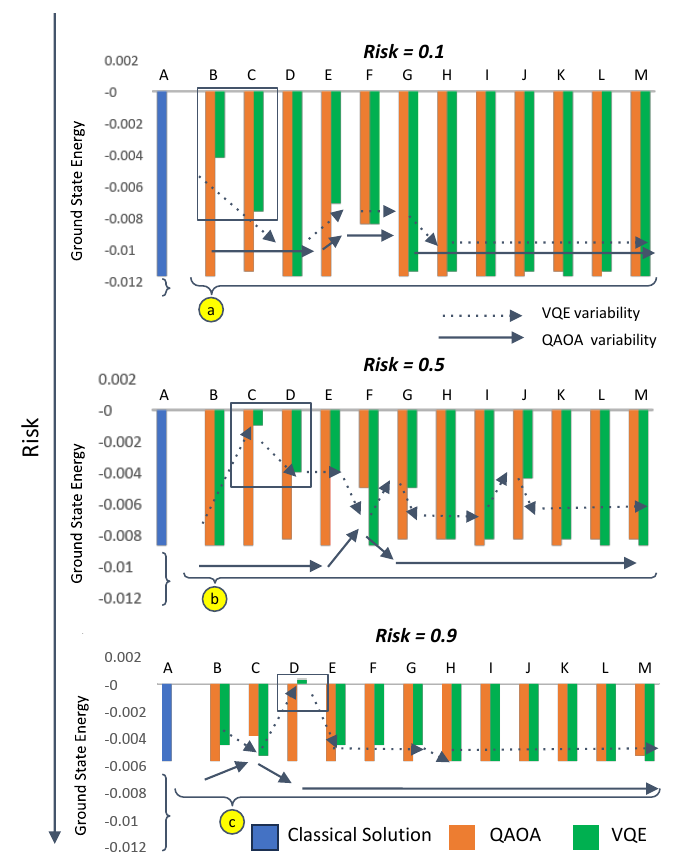}
\centering
\caption{Analysis of ground state energy for different quantum circuit configurations and risk factors, solved using the classical solution, VQE, and QAOA. (1) As indicated by dashed line arrows, the VQE demonstrates very high convergence variability to classical benchmark with testing for mid-level risk margin (\textit{risk = 0.5}), indicating high response to changes induced in the circuit through our experiments across variations from B to M. The QAOA (solid arrow line) exhibits drastically lesser variations across experiments (B to M) for each respective risk value. (3) Comparing QAOA solutions (orange bar) with the classical benchmark (blue bar) across varying experiments and increasing risk, the QAOA is, in most cases, converging to a ground energy value either exactly matching the classical one or closer to it than the VQE. 
(4) The higher the risk, the higher the minimum ground state energy value distribution. as shown in pointers a, b, and c.(5) Certain experiment configurations, highlighted in boxes (B, C, D) in all three figures combined suggest an anomaly in regards to VQE. For these configurations, the VQE results change drastically across all variations, indicating an underlying correlation with parameter values chosen for these configurations. The nature of this correlation can only be evident once investigated further. Still, the consistency of the value change makes it a useful case for a deep dive into the circuit definition for given configurations within the VQE. } 
\label{fig_motivation_results}
\end{figure}

To tackle this problem, we conduct a motivational case study, in which we analyze ground state energy values of different architectural configurations of VQE and QAOA, compared to the classical solution, for low (0.1), middle (0.5), and high (0.9) values of risk factor. The implementation details are discussed in Section~\ref{subsec:exp_setup}. As shown in Fig.~\ref{fig_motivation_results}, some observations can be derived.

\begin{itemize}
    \item The VQE has higher sensitivity to a selected set of quantum parameters (designed with experiment set B to M) than the QAOA, while converging toward the classical benchmark (exact eigensolver solution), tested across three risk values ranging from 0.1 to 0.9. 
    \item By observing the solid and dashed arrow lines in Fig.~\ref{fig_motivation_results}, we see an obvious difference in variability for VQE and QAOA in all three risk value cases. Such variability, which is observable by the changing direction of arrows, is highest in the case of \textit{risk = 0.5} for VQE (dashed arrow lines). On the other hand, the variability for QAOA (solid arrows) has the same pattern, demonstrating consistency and low sensitivity against induced circuit variations and risk values.
    \item The VQE has significant variations across experiments from B to M for different risk margin values. This observation gives us a base to justify a deeper dive into the relationship between the selected quantum parameters and the VQE algorithm to find valuable insights for utilizing the advantage of quantum design for VQE.  
    \item The QAOA exhibits drastically lesser variation over experiments (B to M) for each respective risk value. Additionally, the QAOA is always converging closer or precisely to the classical benchmark solution across experiment variations (B to M) and risk variations.
    \item A holistic view of these three experiments shows that the higher the risk, the higher the minimum ground state energy value distribution, indicated by pointers \textit{a}, \textit{b}, and \textit{c}. This behavior shows that risk values define the landscape of the ground energy values to be minimized, showing a direct relation between the two. 
    \item Highlighted with rectangles, it is observed that some experiment sets have a drastic effect on the converged value for the VQEs. The three boxes in Fig.~\ref{fig_motivation_results} highlight that the experiment sets B, C, and D, which in later sections will be analyzed in more detail, are composed of some parameters or a certain configuration applied to the VQE that cause a drastic variation from rest of the solutions. This is another potential area of investigation to identify what kind of intuitive or counter-intuitive correlations drive this behavior. 
    \item Across many experiments, configurations B, C, and D consistently have an anomalistic behavior when tested with VQE. This makes it useful to investigate further the reason for this dramatic change in value in context with the configuration used. 
\end{itemize}

\subsection{Novel Contributions}

Motivated by the previous analysis, this work tackles the above-discussed challenge by conducting a systematic study of efficient quantum circuit architectures; see Fig.~\ref{fig_POQA_Methodology}. 
We investigate 12 architectural configurations for each optimization algorithm, varying rotation block, entanglement type, and repetitions. In summary, our contributions are:

\begin{figure}[t]
\centering
\includegraphics[width=.95\linewidth]{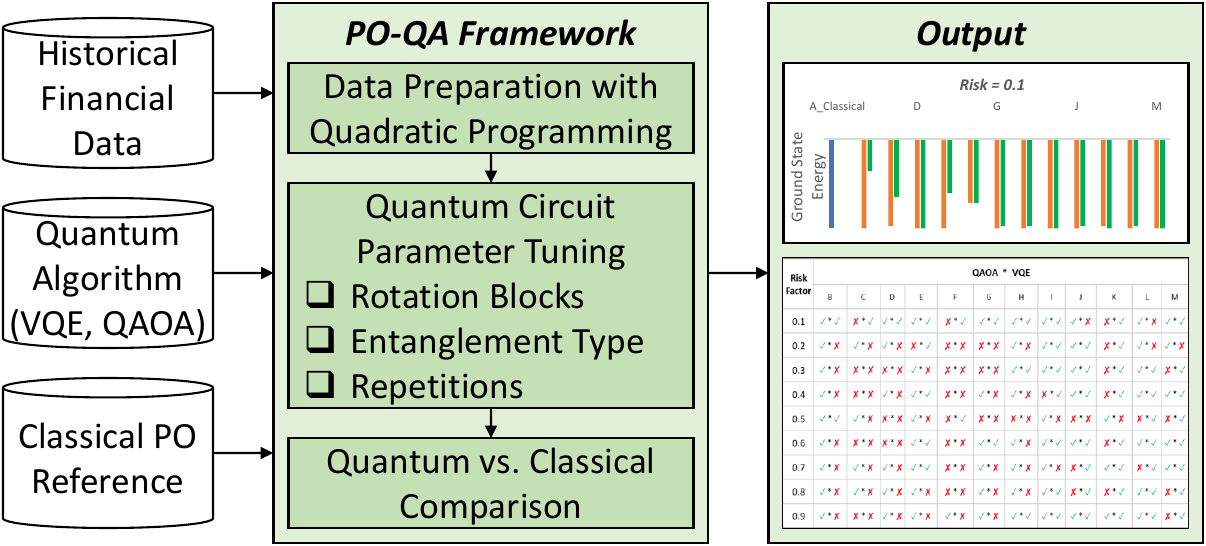}
\caption{Overview of our PO-QA Framework. Given historical financial data, quantum algorithms, and the classical PO reference, the framework performs a quantum circuit parameter tuning and compares the quantum algorithms with the classical reference.}
\label{fig_POQA_Methodology}
\end{figure}

\begin{itemize}
    \item \textbf{PO-QA Framework Design:} After a data preparation stage for quantum algorithms (such as VQE and QAOA) using quadratic programming, our framework tunes the quantum circuit parameters, namely rotation block, entanglement type, and repetitions.
    \item \textbf{Classical vs. Quantum Comparison:} For different risk values, the ground state energy of the classical system has been compared with each quantum circuit architecture. The classical solution is an exact eigensolver used to minimize the ground-energy state landscape. 
    \item \textbf{Key Takeaway:} Our results revealed that the QAOA outperforms the VQE and that interesting correlations between quantum circuit complexity, PO performance, risk, and ground state energy values can be found.
\end{itemize}

\section{Related Work} 
The finance industry is a highly time-critical market, where even micro-seconds can lead to huge gains or losses~\cite{microsecond_market_ieee_spectrum_2012, microsecond_trading_kmad027}. The volatility of this industry presents numerous challenges while seeking accurate and efficient market predictions and optimization processes~\cite{highfrequencyfindatagoodhart1997}. Complex optimization problems form the basis for the potential applicability of high computing infrastructure as an inevitable need for the finance industry. Currently, the resource power and infrastructure are limited, causing hindrances in developing methods for realistic and achievable solutions to problems like large-scale PO. 
Since high-frequency and high-volume data are required to analyze the markets, and numerous external factors drive the market direction, there is a need to develop and adopt compute-intense infrastructures capable of performing tasks beyond classical computers' limitations. Given this nature, financial problems grow exponentially large, making it impossible or intractable for classical computing infrastructures to solve them within useful time frames. The ability to actually make an impact and curate processes for insightful decisions facilitated by machine learning tools and their capabilities lies in the importance of looking beyond current challenges. To lay the base for developing advanced financial predictions on quantum infrastructures, a reasonably efficient direction to take is required~\cite{qmlforfinance_pistoia2021quantum, qmlinfinance_timeseries_2022}. For applicable use cases in the finance industry, encompassing all the factors and variables within the classical computation system becomes an intractable optimization problem. We need machines with greater computation speed and information processing capacity to hold and process numerous factors to produce a generalized result. In this regard, quantum computing methods and infrastructures are considered disruptive technological advancements for the financial industry, not only in long-term, but in use-cases for the near short-term solutions and transformation as well within the NISQ era of quantum computing~\cite{qmlforfinance_pistoia2021quantum}.

Recent works have tackled financial problems in finance. The work of time series forcasting~\cite{qmlinfinance_timeseries_2022} shows that Parametrized Quantum Circuits with simulated quantum forward propagation perform well in the presence of high noise. 
The work on financial forecasting via Quantum Machine Learning~\cite{qmlforecastingfinancethakkar2024} matches the classical performance with significantly fewer parameters.
The works on financial fraud detection using Quantum Graph Neural Networks (GNNs)~\cite{qmlfinfrauddetectioninnan2024} and Quantum Federated Neural Networks (QFNNs)~\cite{innan2024qfnnffdquantumfederatedneural} outperform classical ML methods.
Moreover, a survey of quantum computing for finance~\cite{qcfinanceherman2022survey} presents PO implementations on real hardware using reverse quantum annealing and trapped ion devices, while additionally implementing risk analysis on a transom device. However, the configuration of VQE and QAOA's quantum circuits is unclear. \textit{In this work, we conduct a systematic analysis to identify efficient quantum parameter configurations for PO.}

\section{PO-QA Framework}

Motivated by the previous analysis aimed at studying the effects of quantum parameters tuning on algorithms for an image classification task~\cite{zaman2024_quantumhyperparameterstudy}, in this paper, we focus the analysis on the quantum algorithms for PO. The PO-QA framework builds a foundational base to use and understand quantum financial problem solving, an obvious need for the industry with the momentum that quantum technology has gained. Towards this, in this paper, we present a framework for a systematic investigation of the quantum parameters for PO problems. 

As shown in Figure~\ref{fig_POQA_Methodology_02}, our PO-QA framework is composed of:
\begin{figure*}[t]
\centering
\includegraphics[width=.95\linewidth]{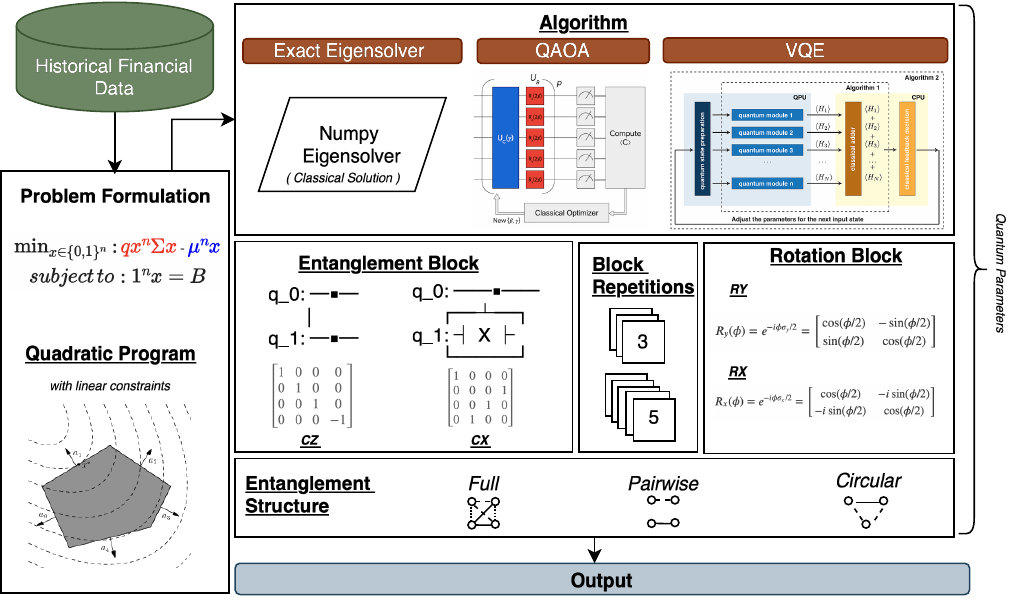}
\caption{Overview of our PO-QA framework's functionality. The problem is formulated through quadratic programming. The exact eignensolver, QAOA, and VQE are used. The quantum circuit has been analyzed by varying the entanglement block, block repetitions, rotation block, and entanglement structure.}
\label{fig_POQA_Methodology_02}
\end{figure*}

\begin{enumerate}
    \item \textbf{Data Preparation:} The financial market data collection required for the given PO problem (and its given constraints, such as number of assets, budget, and risk factor) needs to be converted into quadratic programs with linear constraints for mathematical representation in the quantum system. To process the PO problem, $\mu$ and $\sigma$ are the required inputs from the historical financial data, where $\mu$ is the periodic mean-based expected return per asset and $\sigma$ is the periodic covariance over every portfolio asset.
    \item \textbf{Quantum Circuit Exploration:} Limited variations for each parameter have been selected for the initial development of the framework to restrict the experiment volume to achievable counts. For both VQE and QAOA, these parameters are the \textbf{\textit{rotation blocks}}, i.e., quantum gates that specify the qubit's rotation axis. The gates rx and ry have been used as variations for this parameter. The \textbf{\textit{block repetitions}} and \textbf{\textit{entanglement type}}, which comprises both the \textit{block type} and its \textit{structure (full, pairwise, or circular)} are also varied. The block type refers to the entanglement block, i.e., the gates used in the entanglement layer. These can be specified just like rotation blocks. Additionally, we also tweaked the entanglement structure with the following variations:
    
    \begin{itemize}
        \item \textit{full:} It is an entanglement structure where each qubit is entangled with all the others.
        \item \textit{circular:} In this case, it is linear entanglement but with an additional entanglement of the first and last qubit before the linear part.
        \item \textit{pair-wise:} The case where each qubit is paired with one single qubit exclusively.
    \end{itemize}
    
    The various quantum circuit architectures under investigation form the design space.
    \item \textbf{Quantum vs. Classical Comparison:} The classical solution, implemented with exact eigensolver, outputs the state-of-the-art prediction of the minimum ground state energy of the system. Each quantum circuit architecture of VQE and QAOA predicts the portfolio configuration with a certain ground state energy value, which may or may not be the same as the classical prediction. The comparisons between these values, for different risk factors, enable a systematic evaluation of efficient quantum circuit architectures. \textit{Note that the comparisons are made w.r.t. the ground state prediction obtained by the classical system, which is considered the reference value. In this way, it is possible to identify what parameters of the quantum circuit architectures are suitable for solving this PO problem.}
\end{enumerate}

\section{Evaluating our PO-QA Framework}

\subsection{Experimental Setup}
\label{subsec:exp_setup}

In this work, we use historical financial data of 8 stock values from the Wikipedia DataProvider 
interface provided by Qiskit, in a time period of 6 months from 1st July 2016 to 31st December 2016. The other experiment specifications are reported in Table~\ref{tab:exp_setup}.

\begin{table}[ht]
\centering
\caption{Experiment Specifications.}
\label{tab:exp_setup}
\begin{adjustbox}{max width=.9\linewidth}
\begin{tabular}{c|c}
\textbf{Parameter} & \textbf{Specification} \\ \toprule
Software Framework & \verb|Qiskit(QK)| \\ \midrule
 Algorithm & \verb| Exact Solver(classical), VQE, QAOA|   \\ \midrule
Rotation Block &\ \verb|ry, rx|\\ \midrule
Entanglement Block & \verb|cz, cx| \\ \midrule
 Entanglement Structure & \verb|full, pairwise, circular|\\ \midrule
Block Repetitions (circuit depth) &  3 and 5 \\ \midrule
Risk Factor & \verb|0.1|, \verb|0.2|, \verb|0.3|, \verb|0.4|, \verb|0.5|, \verb|0.6|, \verb|0.7|, \verb|0.8|, \verb|0.9| \\ \midrule
Asset Count, Budget & \verb|8|, \verb|assets/2|\\ 
\midrule
Assets & \verb|TSLA|, \verb|AMZN|, \verb|GOOG|,\verb|AAPL|, \verb|FSLR|, \verb|SPWR|, \verb|ARRY|, \verb|ENPH| \\ 
\midrule
\end{tabular}%
\end{adjustbox}
\end{table}

\begin{figure*}[ht]
\includegraphics[width=\linewidth]{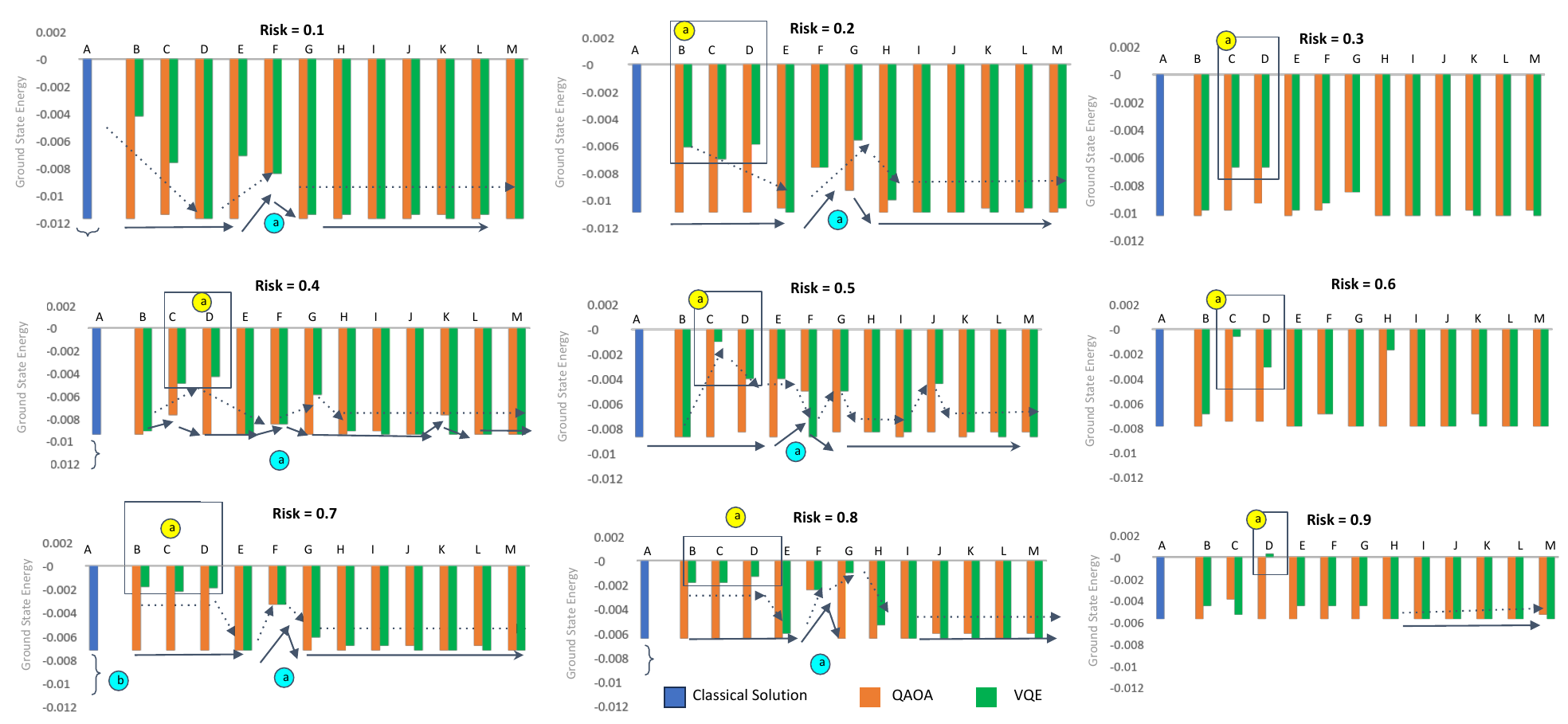}
\centering
\caption{Variation of the predicted ground state energy level, for different quantum circuit architecture configurations of VQE (green) and QAOA (orange) solutions, w.r.t. the classical solutions (blue), varying the risk factor. Here, the solid arrow line indicates QAOA variations across experiments and risk, and the dashed arrows indicate VQE variability. Blue markers with label \textit{a} signify the resistance of QAOA to induced parameter tweaking (low sensitivity, mostly the same pattern), but it also highlights that at blue marker \textit{a}, the configuration usually F and sometimes additionally G cause a sudden increase in ground-state energy value. The consistency of this behavior of QAOA with configuration F allows us to reason the correlation and understand better the underlying cause to find useful insights. The pink label \textit{a} indicates that out of all experimental results for VQE, a risk margin of 0.5 causes most instabilities and variations across configurations (B to M). As highlighted by blue marker \textit{b}, for risk=0.7 and at similar places for results of risk=0.1, risk=0.4, risk=0.8, and risk=0.9, we observe that the higher the risk, the higher the minimum ground-state energy value is returned.}
\label{fig_bars_values}
\end{figure*}

\subsection{Results: Classical vs. Quantum Comparison}

\begin{figure}[t]
 \centering
\includegraphics[width=.9\linewidth]{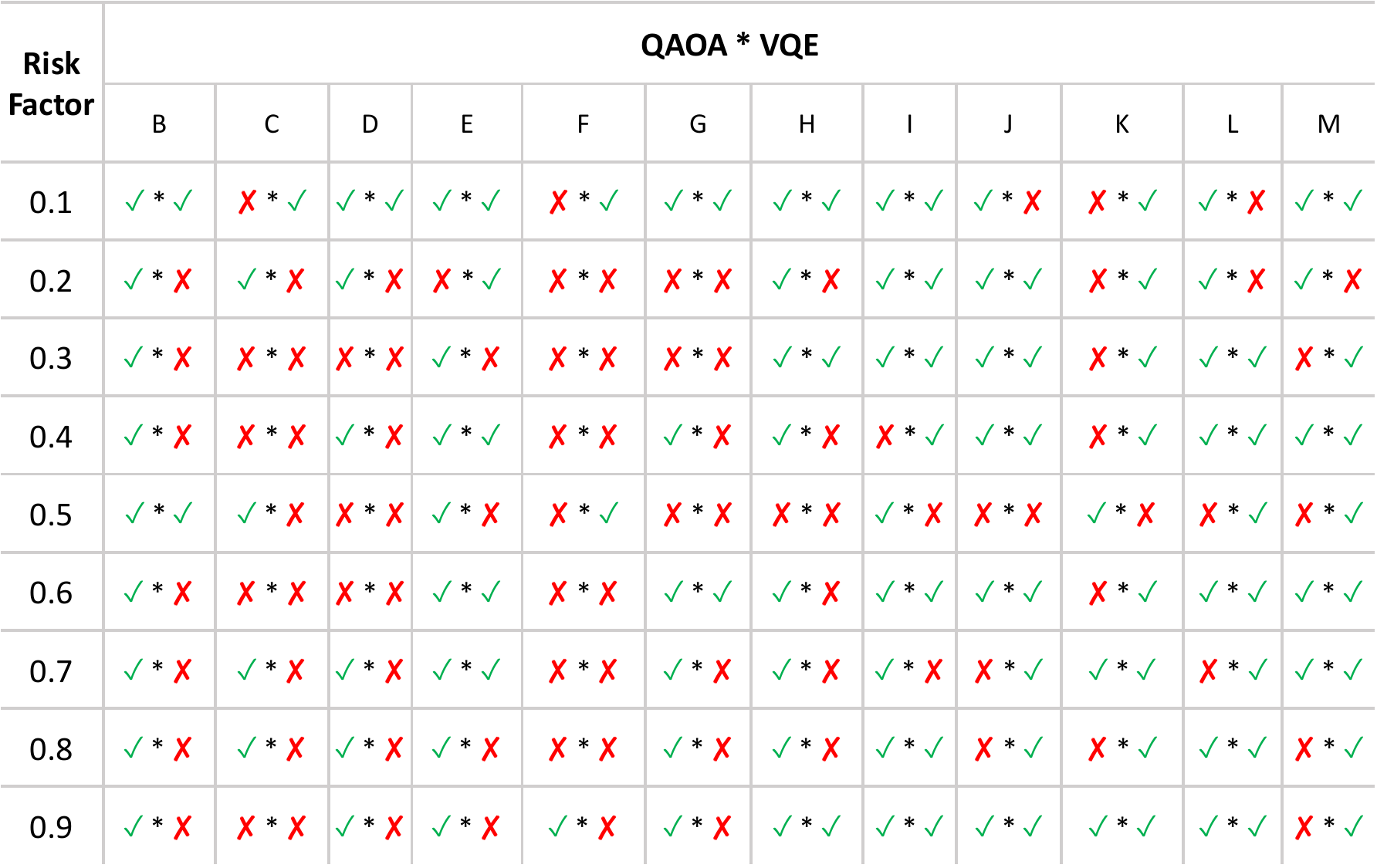}
 \caption{Classical vs. Quantum comparison: for each risk factor, the table reports whether the minimum ground state predicted by the QAOA and VQE systems matches the prediction of the classical system. The green tick identifies a matched case, while the red cross indicates that the resulting optimal portfolio asset configuration differs from that of the classical solution. Legend: B=(full, ry, cz, 3), C=(pairwise, ry, cz, 3), D=(circular, ry, cz, 3), E=(full, ry, cz, 5), F=(pairwise, ry, cz, 5), G=(circular, ry, cz, 5), H=(full, rx, cx, 3), I=(pairwise, rx, cx, 3), J=(circular, rx, cx, 3), K=(full, rx, cx, 3), L=(pairwise, rx, cx, 3), M=(circular, rx, cx, 5).}
 \label{tab_results_classical_vs_quantum}
 \end{figure}

For each quantum algorithm (VQE and QAOA), 12 different quantum circuit architecture variations (labeled with letters from B to M) are tested. Figures~\ref{fig_bars_values} and~\ref{tab_results_classical_vs_quantum} show the comparison of each quantum circuit architecture variation, for different values of the risk factor, ranging from 0.1 to 0.9 along with the classical solution. Fig.~\ref{fig_bars_values} visualize the differences of the ground state energy values among QAOA and VQE variations against the classical solution as a benchmark, while Fig.~\ref{tab_results_classical_vs_quantum} reports whether each quantum configuration matches or converges to the classical solution in terms of minimum ground state energy (i.e., predicted optimal portfolio).

From the results, we can observe that the QAOA is mostly insensitive to the quantum circuit variations we have performed in the context of PO, whereas the VQE demonstrates extremely high sensitivity towards indices variations. This observation has been consistent throughout most experiments conducted with various risk factors. Hence, it proves to be a valid ground for assuming a deterministic relationship between the VQE and these parameters, causing regular variations. A deep dive into uncovering this relationship can help us discover insights into the design and working of the VQE. On the other hand, the QAOA has low sensitivity. However, there are instances of repeated behavior across similar experiments. 

\begin{figure}[ht]
\centering
\includegraphics[width=.95\linewidth]{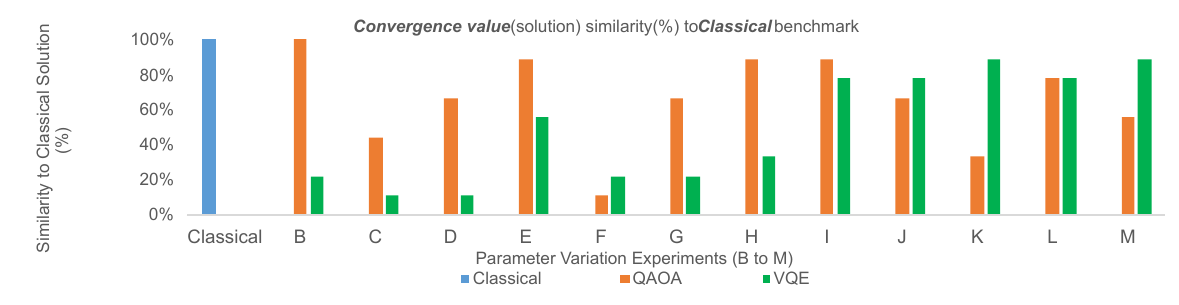}
\vspace*{-5pt}
\caption{Percentage comparison of QAOA and VQE for each experiment configuration to classical benchmark (exact eigensolver) solution.}
\label{fig_po_qa_matchcases}
\end{figure}

Overall, we can observe that the QAOA tends to perform better than the VQE since more QAOA architectures match with the baseline classical optimal portfolio as observable in Fig. \ref{fig_po_qa_matchcases}.  This can be due to the fact that the QAOA is a more advanced algorithm built upon the principles of the VQE. The QAOA's additional constraints and improved problem construction show that this algorithm finds the minimum energy easier than the VQE, since only the QAOA with quantum circuit architecture F performs poorly. On the other hand, the results of the VQE consistently vary based on the quantum circuit architecture. For instance, we can observe that 5 repetitions outperform 3. We can derive that the circuit depth is directly correlated with the performance. 

We can also observe that experiments with low risk achieve a lower ground state energy, while increasing the risk progressively increases the energy value. This is due to the correlation between risk factor and ground energy level for optimal portfolios.

Fig.~\ref{fig_po_qa_matchcases} clearly highlights that the QAOA has a higher matching rate with the exact eigensolver, which is being used as a classical benchmark. The percentage shows the share of experiments that resulted in exactly the same solution as the classical solver (A) for the respective risk value (0.1 to 0.9) and experiment configuration (B to M) type.

\textit{In summary, our systematic analysis conducted with the PO-FA framework unveiled the following key takeaways:}

\begin{itemize}
    \item The QAOA outperforms the VQE because it solves the PO problem using quantum circuit architectures that are less complex.
    \item The higher the quantum circuit architecture complexity (i.e., depth), the better the VQE algorithm performs.
    \item The higher the risk value, the higher the ground state energy.
\end{itemize}

\section{Discussion}

The novelty of our work lies in its aim to address the need for basic demonstrative quantum framework implementation, which facilitates research and circuit design development for financial solutions through the quantum machine learning lens.


 Collectively, critical aspects of the finance industry problems and solutions depend on quick processing, critical timing, and transaction safety. The need for particular accuracy, given the high-risk stakes, involves the need for information to consume the maximum available market data and to process and predict complementing insights. Our effort is driven by the need to provide a basic setup and direction toward exploring financial solutions using quantum algorithms to analyze the solutions from the lens of catering to unmet financial market processing requirements compared to classical solutions. 

 The following are the key aspects that make our work useful for the research and design community.

\subsection{Real-Market Data Analysis}
 The true value of our work lies in presenting these solutions on \textbf{\textit{real-market data for a substantial time series (6 months).}} The data volume in PO is ever-increasing. The closer we map the problem to the real-world setting, the more data and constraints have to be catered for. This drastic increase in the amount of information required for computing and processing usually turns slightly complex PO tasks into intractable complex problems on classical computers. Our aim in showcasing solutions on the quantum infrastructure is to convince the industry of its potential as a beneficial technological disruption to enable the infrastructure to \textbf{\textit{handle the constantly growing information and processing capacity for complex but more realistic PO problems and solutions,}} upholding potential for rich portfolio diversification which otherwise is a challenging task.
 
\subsection{High experiment volume and variation}
The intentional decision for deriving results on a \textbf{\textit{significantly large and varying experiment volume}} incorporated with space within the framework for further expansion towards exploratory directions and variations is an absolutely necessary and desired step towards comprehending the underlying interactions in PO processing to uncover effective relationships between particular processes and its applicability, advantage, and usefulness for implementation on quantum infrastructures. This allows us to uncover and better understand the optimal decisions made by the PO solution implemented on the quantum infrastructure, compared to which results optimally on classical counterparts. Experimenting variations of quantum algorithms like VQE and QAOA with diverse experiment sets for each, varying over tunable quantum parameters (entanglement block and type, depth of circuit, etc.), allows to test solutions to the optimization problem with design margin for scalability to incorporate further variations within the quantum parameters.
\subsection{Scalable framework}
The framework is designed to easily be scaled up to a larger set of experiment variations. Testing out additional optimization algorithms opens up the space for algorithmic design creativity to build upon these implementations and discover or formulate more solutions to the PO problem, outperforming the current ones or uncovering underlying relationships among fine-tuned parameters and their effects on the convergence to the classical benchmark solution. Newer methods and strategies for performance metrics and design can also be an aspect of potential future use cases.

The scalability of this framework is an important aspect. We have showcased basic experiments and studies following along the same lines and using these implementations and analyses as building blocks that can be utilized to build and grow the understanding of quantum-based solutions for finance applications and operations. Given the information handling capacity of Quantum Computing, aiming for a given problem's context-aware circuit design and implementation is one of the useful and impactful directions. Efficiency is very subjective in the finance domain, despite having similar themes or standards driving the performance, like fast latency and risk trade-off.
\subsection{Larger target audience}
The inclusion of a diverse target audience is what drives simple approaches to results and analysis. Enables hands-on exposure to foundational framework developments for Quantum Finance for expansion into further research potentials. An added value of this kind of work is to \textbf{\textit{facilitate the finance community into familiarizing themselves with a highly technical quantum infrastructure available}} at the time with simple, straightforward, and interpretable demonstrations of relevant work in the financial industry.
\subsection{Reproducible base experiments for deeper study towards quantum finance }
The above-discussed aspects lay the foundation for quantum systems for finance. Functioning on the principle of outsourcing some complex parts to QC and getting accurate results, the workload distribution is optimized between classical and quantum infrastructures.



\section{Conclusion}

In this work, we have proposed PO-QA, a framework for exploring quantum circuit parameters of VQE and QAOA for PO problems. For each quantum circuit architecture and various ranges of risk factors, it compares the predicted ground state energy w.r.t. the estimated classical solution. Our findings reveal interesting correlations between quantum parameters that pave the way toward efficiently solving quantum PO problems. We consider this work important to build forward in the direction of implementing and exploring quantum algorithms for finance in research and exploratory purposes to enable design experimentation. We see potential as we could identify effective and interpretable results, correlations, and differences in varying parameters, uncovering underlying patterns over changing risk values and convergence to classical solutions.

Another important aspect and potential of future work building upon this framework is to showcase a hard optimization problem in the finance industry that actually solves a previously unsolvable task in the financial context, which provides value and motivation for the need for quantum. Moreover, a comparison outperforming a similar classical setup would showcase the ability to fine-tune quantum circuits, which is particularly suitable for the quantum finance industry. The future outlook lies in corporate banking, an inevitable domain with high monetary value at stake, limited resources, and strict constraints and policies that make it stand out as the most influential and important stakeholder.

Quantum Machine Learning for finance presents itself as an opportunity to face the current financial industry's computing challenges. It caters to quicker computations driven by the acclaimed strength of a quantum computer capable of faster computing. This advantage facilitates the fast-moving markets that require quick computing or fast-paced computing to keep up with real-world information.









\section*{Acknowledgments}
    This work was supported in part by the NYUAD Center for Quantum and Topological Systems (CQTS), funded by Tamkeen under the NYUAD Research Institute grant CG008.

\bibliographystyle{IEEEtran}  
\bibliography{main.bib}

\end{document}